\documentclass[aps,pra,reprint,superscriptaddress,showpacs]{revtex4-1}
\usepackage[utf8]{inputenc}
\usepackage{graphicx}
\usepackage[breaklinks=true,colorlinks=true,linkcolor=blue,urlcolor=blue,citecolor=blue]{hyperref}
\begin{document}

%\preprint{}

\title{Near-K-edge single, double, and triple photoionization of C$^+$ ions}

% Please use the appropriate macro for each type of information

\author{A. M\"{u}ller}
\email[]{Alfred.Mueller@iamp.physik.uni-giessen.de}
\affiliation{Institut f\"{u}r Atom- und Molek\"{u}lphysik, Justus-Liebig-Universit\"{a}t Gie{\ss}en, Giessen, Germany}
%\thanks{}
\author{A. Borovik Jr.}
%\affiliation{Institut f\"{u}r Atom- und Molek\"{u}lphysik, Justus-Liebig-Universit\"{a}t Giessen, Germany}
\affiliation{I. Physikalisches Institut, Justus-Liebig-Universit\"{a}t Gie{\ss}en,  Giessen, Germany}
\author{T.~Buhr}
%\affiliation{Institut f\"{u}r Atom- und Molek\"{u}lphysik, Justus-Liebig-Universit\"{a}t Giessen, Germany}
\affiliation{I. Physikalisches Institut, Justus-Liebig-Universit\"{a}t Gie{\ss}en, Giessen, Germany}
%\affiliation{Physikalisch-Technische Bundesanstalt,  Braunschweig, Germany}
\author{J. Hellhund}
\affiliation{Institut f\"{u}r Atom- und Molek\"{u}lphysik, Justus-Liebig-Universit\"{a}t Gie{\ss}en, Giessen, Germany}
\author{K. Holste}
%\affiliation{Institut f\"{u}r Atom- und Molek\"{u}lphysik, Justus-Liebig-Universit\"{a}t Giessen, Germany}
\affiliation{I. Physikalisches Institut, Justus-Liebig-Universit\"{a}t Gie{\ss}en, Giessen, Germany}
\author{A.~L.~D.~Kilcoyne}
\affiliation{Advanced Light Source, Lawrence Berkeley National Laboratory, Berkeley, California, USA}
\author{S. Klumpp}
%\affiliation{Institut f\"{u}r Experimentalphysik, Universit\"{a}t Hamburg, Hamburg, Germany}
\affiliation{DESY Photon Science, FS-FLASH-D, Hamburg, Germany}
\author{M. Martins}
\affiliation{Institut f\"{u}r Experimentalphysik, Universit\"{a}t Hamburg, Hamburg, Germany}
\author{S.~Ricz}
%\affiliation{Institut f\"{u}r Atom- und Molek\"{u}lphysik, Justus-Liebig-Universit\"{a}t Giessen, Germany}
\affiliation{Institute for Nuclear Research, Hungarian Academy of Sciences, Debrecen,  Hungary}
\author{J. Viefhaus}
\altaffiliation{Present address: Helmholtz-Zentrum Berlin, Department Optics and Beamlines, Berlin, Germany}
\affiliation{DESY Photon Science, FS-PE, Hamburg, Germany}
\author{ S. Schippers}
%\affiliation{Institut f\"{u}r Atom- und Molek\"{u}lphysik, Justus-Liebig-Universit\"{a}t Giessen, Germany}
\affiliation{I. Physikalisches Institut, Justus-Liebig-Universit\"{a}t Gie{\ss}en, Giessen, Germany}


\date{\today}

\begin{abstract}
Single, double, and triple ionization of the C$^{+}$ ion by a single photon  have been investigated in the energy range 286 to 326~eV around the K-shell single-ionization threshold at an unprecedented level of detail. At energy resolutions as low as 12~meV,  corresponding to a resolving power of  24\,000, natural linewidths of the most prominent resonances could be determined. From the measurement of absolute cross sections, oscillator strengths, Einstein coefficients, multi-electron Auger decay rates  and other transition parameters of the main K-shell excitation and decay processes are derived. The cross sections are compared to results of previous theoretical calculations. Mixed levels of agreement are found despite the relatively simple atomic structure of the C$^+$ ion with only 5 electrons. This paper is a follow-up of a previous Letter [M\"{u}ller {\textit{et al.}}, Phys. Rev. Lett. {\textbf{114}}, 013002 (2015)].
\end{abstract}



%\keywords{hollow ions, resonances, interference, double-Auger decay, excitation-autoionization}

\maketitle

%\section{Introduction}

\section{Introduction}

Carbon is the second most abundant heavy element (with atomic numbers $Z>2$) in the universe next to oxygen. It is prominently present in all astrophysical environments. Moreover, it is the prime constituent of organic chemistry and the building block of life on Earth. In both, collisionally ionized and photoionized gases, carbon atoms occur singly or multiply charged and the ions can serve as probes for the state of the plasma in which they are formed, be it of astrophysical or terrestrial origin. Important diagnostic techniques rely on the soft-x-ray spectroscopy of carbon ions especially near the K-edge~\cite{Hasoglu2010}.

Similar spectroscopic techniques based on the production and observation of the decay of K-vacancy states are employed to characterize the chemical environment of carbon atoms in molecules, clusters, and solids. When a K-shell electron is removed from a neutral carbon atom the resulting ion is most likely in one of the C$^{+}(1s2s^22p^2~^2{\rm S},^2\!{\rm P},^2\!{\rm D})$ states which then decay by emitting photons or electrons with characteristic energies. Identical terms can be excited from the ground state of C$^{+}$ ions by irradiation with photons of the same characteristic energies~\cite{Schlachter2004a}. Resonance parameters and details of the transitions such as the natural widths, the oscillator strengths and the exact wavelengths can be determined in photoionization experiments with C$^{+}$ ions.

An important aspect of K-shell excited C$^{+}(1s2s^22p^2~^2{\rm S},^2\!{\rm P},^2\!{\rm D})$ states is their unique electronic configuration with four electrons in the L-shell. Four electrons residing above a K-shell vacancy is the minimum number required for observing triple-Auger decay in which one of the four electrons falls into the K vacancy while all three other electrons are ejected into the continuum in a correlated four-electron process. This process has been unambiguously identified in our preceding Letter~\cite{Mueller2015a}.

Photoprocesses involving C$^+$ ions have been studied in previous experiments. Photoabsorption by boronlike C$^{+}$ ions was investigated by Jannitti {\textit{et al.}}~\cite{Jannitti1993a} with a technique using two laser-produced plasmas.
Photoionization of  C$^{+}$ ions has already been addressed in several merged-beam experiments. The valence-electron energy range has been studied by Kjeldsen {\textit{et al.}}~\cite{Kjeldsen1999a,Kjeldsen2001b}, while Schlachter {\textit{et al.}}~\cite{Schlachter2004a} explored the $1s \to 2p$ excitation resonances. In both cases absolute single-ionization cross sections were determined. Theoretical calculations for K-shell photoabsorption by C$^{+}$ ions have produced cross sections using R-matrix techniques~\cite{Schlachter2004a,Wang2007,Hasoglu2010} and, indirectly, by the multi-configuration Dirac-Fock method~\cite{Shi2009a}.

Measurements for the boronlike sequence of ions were recently extended to N$^{2+}$~\cite{Gharaibeh2014} and O$^{3+}$~\cite{McLaughlin2014} by using the photon-ion merged-beam technique. Resonant K-shell excitation of boronlike Fe$^{21+}(1s^22s^22p~^2{\textrm P}_{1/2})$ was observed by detecting fluorescence photons and photoions~\cite{Rudolph2013,Steinbruegge2015} in experiments employing an electron beam ion trap. Deep-core photoexcitation of ions has recently been reviewed by M\"{u}ller~\cite{Mueller2015c}.

Due to the low density of ions in a beam, the presence of high detector backgrounds and limited photon flux available from synchrotron radiation sources,  photoionization experiments involving K-vacancy production had been limited until recently to the strongest resonance features in the spectrum. Schlachter {\textit{et al.}}~\cite{Schlachter2004a} observed only $1s \to 2p$ transitions in C$^{+}$ at resolving powers 2400 to 5800. Gharaibeh {\textit{et al.}}~\cite{Gharaibeh2014} measured the strongest $1s \to 2p$ and $1s \to 3p$ transitions in N$^{2+}$ at resolving powers E/$\Delta$E 4500 to 13\,500. McLaughlin {\textit{et al.}}~\cite{McLaughlin2014} investigated the same transitions in O$^{3+}$ at resolving powers E/$\Delta$E 3200 to 5000. The ion-trap experiments involving K-vacancy production in B-like Fe$^{21+}$ ions were restricted to the observation of an unresolved blend of Fe$^{21+}(1s2s^22p^2~^2{\textrm P}_{1/2})$ and Fe$^{21+}(1s2s^22p^2~^2{\textrm D}_{3/2})$ levels in the energy range 6583 - 6589~eV. The associated resonances were not directly measured on an absolute cross-section scale. The energy resolution in the particular measurement on Fe$^{21+}$ was 1~eV corresponding to a resolving power of about 6600.

All those previous experiments were restricted to the single-ionization channel and few isolated resonances. Since then, significant experimental progress has been achieved in that the energy ranges of experiments have been extended to the region beyond the K edge and multiple ionization up to the removal of three electrons from the initial ion was investigated for the lighter elements up to atomic number $Z = 10$~\cite{Mueller2015a,Bizau2015,Mueller2015d,Schippers2016a,Mueller2017}. L-shell excitation and ionization in Fe$^+$ and M-shell vacancy production in singly and multiply charged xenon ions have been investigated by observing up to 6-fold net ionization of the parent ion~\cite{Schippers2017,Schippers2014,Schippers2015a}.

Here we report absolute cross-section measurements for photoionization of C$^{+}$ near the K edge. The experiments were performed at the photon-ion merged-beam setup PIPE (\underline{p}hoton-\underline{i}on spectrometer at \underline{PE}TRA III)~\cite{Schippers2014} using monochromatized undulator radiation from the PETRA~III synchrotron light source in Hamburg. Single and double ionization were investigated covering the whole range of $1s \to np$  ($n = 2, 3, 4,...,\infty$) one-electron and $1s2s \to 2pnp$  ($n = 2, 3, 4,...,\infty$) two-electron excitations occurring well beyond the K-shell ionization threshold. For the strongest $1s \to 2p$ resonances and the immediate K-edge region triple ionization was also observed. With the extended capabilities of the PIPE experiment, structures and processes in ionized atoms, molecules and clusters can be observed which have not been previously accessible to experiments.

The present paper provides details of experiments and additional results from our previous work~\cite{Mueller2015a}. Emphasis in the previous publication was on the discovery and unambiguous demonstration of an elusive four-electron Auger process, the triple-Auger decay, in which three electrons are ejected in a single event while a fourth electron falls into a K-shell vacancy. The present paper focuses on the determination of absolute cross sections at very high energy resolution and the information that can be derived from the combined results of absolute measurements of partial cross sections for different final channels and the natural widths of resonance lines.

This presentation is structured as follows. After this introduction, a brief overview of the experiment is provided with details specific to the present measurements. The experimental results are shown in detail. Photoabsorption cross sections for C$^+$ ions and decay parameters of core excited states are derived from the measurements and compared with the results of theoretical calculations as well as with measurements available in the literature. After a summary and acknowledgements, an appendix describes the formalism and procedure for the extraction of transition parameters from absolute photoionization cross sections.

\section{Experiment}

The experimental arrangement and procedures have been described in detail previously~\cite{Schippers2014,Mueller2017}. In short, C$^{+}$ ions were produced for the present measurements in an electron-cyclotron-resonance (ECR) ion source. The ions were accelerated to 6~keV, magnetically analyzed to obtain an isotopically pure beam which was then transported to the interaction region, collimated and merged with the photon beam available at beamline P04~\cite{Viefhaus2013} of PETRA~III. The product ions were separated from the parent ion beam by a dipole magnet inside which the primary beam was collected in a large Faraday cup. The photoionized ions were  passed through a spherical 180-degree out-of-plane deflector to suppress background from stray electrons, photons and ions and then entered a single-particle detector with near-100\% detection efficiency. The high brightness and flux of the photon beam ($2\times10^{11}$~s$^{-1}$ at 288~eV energy and 12~meV bandwidth) permitted tight collimation of the ion beam (1.5 nA) and a significantly improved spatial overlap with the photon beam compared to earlier experiments. Form factors~\cite{Phaneuf1999,Schippers2014,Mueller2017}  in the cross-section measurements were between 3460 and 4800~cm$^{-1}$; optimized beam tuning resulted even in 9900~cm$^{-1}$. In the preceding Letter~\cite{Mueller2015a} the single-interaction condition (strictly only one photon absorbed and no other interactions) for the measurement of the relatively small triple-ionization cross section has been discussed in detail and unquestionably verified.

The photon flux was measured with a calibrated photodiode. The photon energy scale was calibrated with an uncertainty of better than $\pm$30~meV by remeasuring known photoionization resonances in  C$^{3+}$~\cite{Mueller2009a}. Doppler shifts due to the ion velocity directed opposite to the incoming photons were corrected for, the correction factor of the laboratory photon energy being approximately 1.001.  The systematic total uncertainty of the measured cross section for single and double ionization is $\pm$15\%~\cite{Schippers2014} to which statistical uncertainties have to be added. In the case of triple ionization  the cross section was derived from ratios between consecutive recordings of the spectra for single, double, and triple ionization with the experimental conditions unchanged and by normalizing to the absolute data for single and double ionization. A conservative estimate  of the uncertainty of the extremely small absolute triple-ionization cross section is $\pm$50\%.

\section{Results}

 \begin{figure*}
\includegraphics[width=16.5cm]{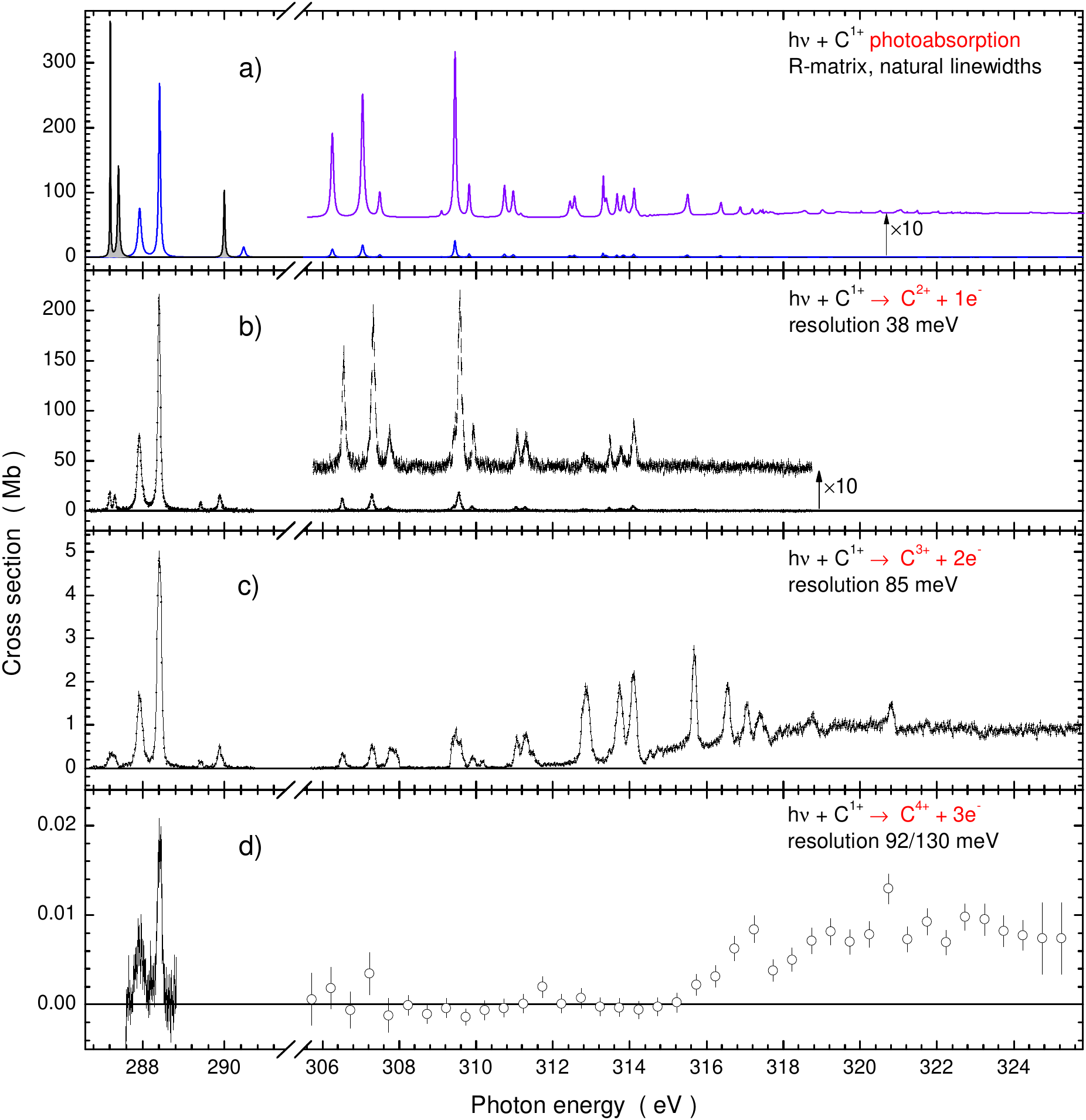}
\caption{\label{Fig:overview} (color online) Overview of the cross sections for photoionization of C$^{+}$ ions addressed in this paper. Panel a) shows theoretical cross sections~\cite{Hasoglu2010} for photoabsorption by long-lived excited C$^{+}(1s^2 2s 2p^2~^4{\rm P}$) and ground-term C$^{+}(1s^2 2s^2 2p~^2{\rm P}$) parent ions. The R-matrix results for the metastable ions are restricted to $ 1s \to 2p$ transitions. They are represented by the black solid line with gray shading. The calculated data for ground-term ions cover the full energy range 287 - 325~eV investigated in the present experiments. They are represented by the lighter (blue) solid line. Multiplication of the ground-term cross section by a factor of 10 produced the (purple) line which is shown with a vertical offset for better visibility. The energy axis has a break between 294 and 305~eV where no resonances are expected.
Panel b) displays experimental results for single photoionization of C$^+$ ions measured at an energy resolution of 38~meV. The cross sections obtained beyond 304~eV are shown a second time after multiplication by a factor of 10. They are displayed with an offset. All experimental cross sections are on an absolute scale. Panel c) shows experimental results for double photoionization of C$^+$ ions measured at an energy resolution of 85~meV. In panel d) cross sections for triple photoionization of C$^+$ ions are displayed. The densely spaced data at around 288~eV were measured at an energy resolution of 92~meV, the data points with statistical error bars beyond 305~eV were taken at 130~meV resolution.
}
\end{figure*}

Figure~\ref{Fig:overview} provides an overview of the investigated cross sections for single, double, and triple ionization of C$^+$ ions by single photons. Theoretical calculations for photoabsorption by ground-term and metastable C$^+$ ions~\cite{Hasoglu2010} are displayed in panel a). For metastable  C$^{+}(1s^2 2s 2p^2~^4{\rm P}$) ions with an excitation energy of about 5.33~eV~\cite{NIST2015} the available calculations are restricted to the energy range where $1s \to 2p$ excitations occur. The cross sections are represented by a black solid line with gray shading. For ground-term C$^{+}(1s^2 2s^2 2p~^2{\rm P}$) ions the lighter (blue) solid line was calculated. Since the cross sections at energies beyond 305~eV are relatively small as compared to the $1s \to 2p$ resonances they were multiplied by a factor 10 and displayed again in panel a) with a vertical offset.

Panel b) shows the experimental cross section for net single ionization of C$^+$ ions by single photons measured at a bandwidth of 38~meV. Like all other cross sections discussed in this paper these data are on an independently absolute scale. The spectrum at energies beyond 305~eV was multiplied by a factor of 10 and displayed again in panel b) with a vertical offset for better visibility of the small resonance peaks. In panel c) the experimental cross section for net double ionization of the C$^+$ ion by a single photon is displayed. It was measured at a bandwidth of 85~meV. The vertical scale is down from panel b) by roughly a factor of 40. Panel d) shows experimental cross sections for net triple ionization of the C$^+$ ion by a single photon. The energy region around 288~eV was scanned in fine steps at a resolution of 92~meV. The isolated cross section data at higher energies were measured at a bandwidth of 130~meV. Unfortunately, the low signal rates were not sufficient for scanning a wide energy range in small steps so that resonance structures expected in the energy range 305 - 325~eV could not be made visible unambiguously. Note that the cross-section scale is down from that of panel c) by another factor of about 200.

\begin{figure*}
\includegraphics[width=13cm]{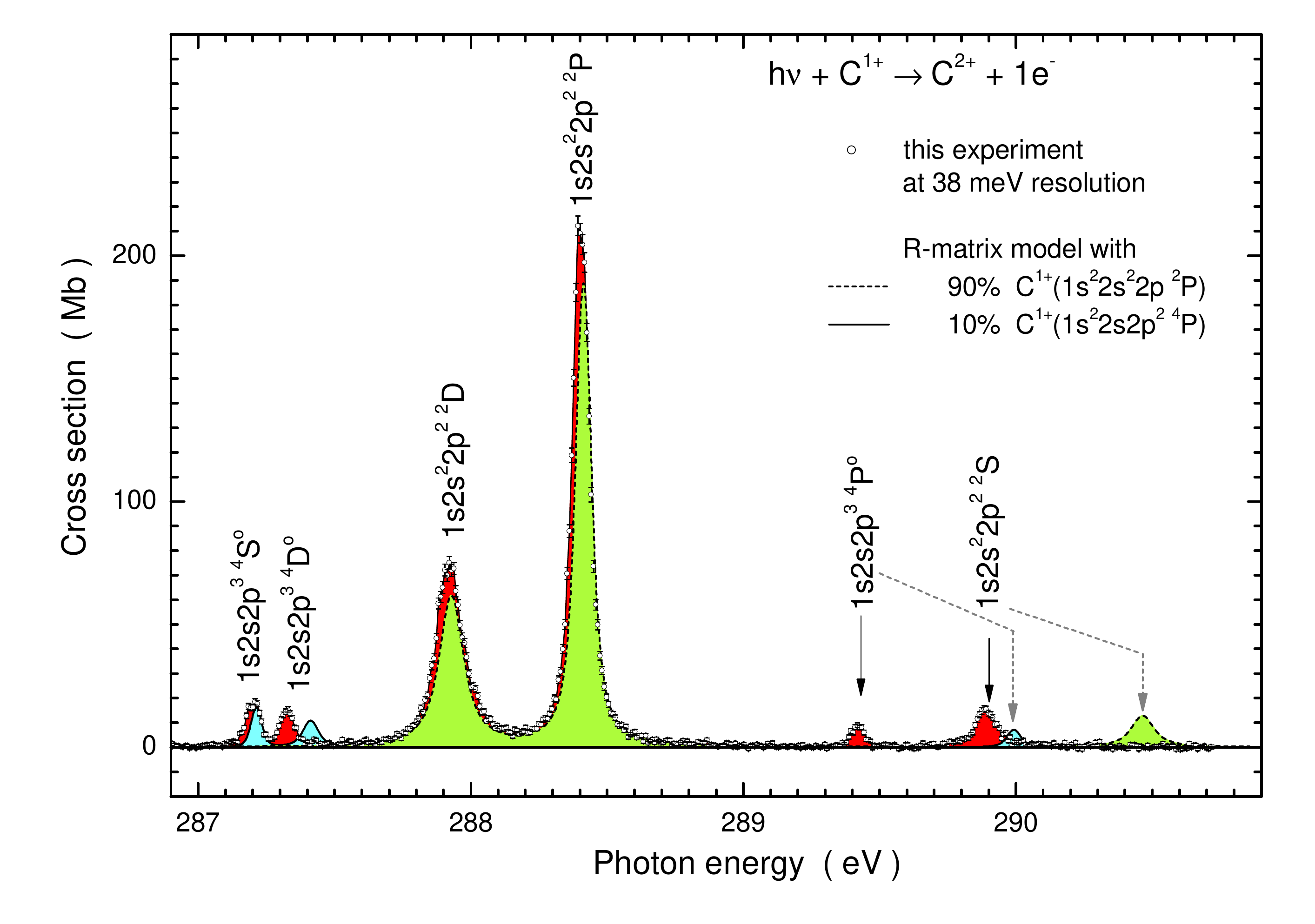}
\caption{\label{Fig:1s2p} (color online) Experimental cross section for single photoionization of C$^{+}$ ions in the region of $1s \to 2p$ excitations at 38~meV resolution compared with results of R-matrix photoabsorption calculations~\cite{Hasoglu2010} convoluted with a 38~meV full-width-at-half-maximum (FWHM) Gaussian distribution function. The terms associated with the individual resonance peaks are indicated. The present measurement is represented by shaded circles with statistical error bars. The area under the experimental cross section is dark shaded (red online). The primary ion beam contained an approximate 10\% fraction of metastable C$^{+}(1s^22s2p^2~^4{\rm P}$) and 90\% ground-term C$^{+}(1s^2 2s^2 2p~^2{\rm P}$) ions. Accordingly, the R-matrix photoabsorption cross sections are a weighted sum of the different parent ion fractions which are individually presented: quartet resonances from metastable parent ions by a solid line with (cyan) shading and doublet resonances from ground-state ions by the dashed black line with (green) shading.  }
\end{figure*}

Figure~\ref{Fig:1s2p} shows the result of a photon-energy scan at 38~meV resolution of the cross section for single photoionization of C$^{+}$ ions by single photons in the energy range 286.9 - 290.9~eV covering all contributions from $1s \to 2p$ excitations. Individual resonance terms are identified on the basis of the discussion provided in the work by Haso$\breve{\rm g}$lu {\textit{et al.}}~\cite{Hasoglu2010}. The assignment of the first two peaks is reversed compared to earlier theoretical work~\cite{Schlachter2004a,Wang2007}. Arguments for the new assignment are provided in the theory paper published by Haso$\breve{\rm g}$lu {\textit{et al.}}~\cite{Hasoglu2010}. The quartet terms arise from metastable C$^{+}(1s^22s2p^2~^4{\rm P})$ parent ions, the doublet terms are populated by excitation of ground-level C$^{+}(1s^22s^22p~^2{\rm P})$ ions. Both, ground-state and metastable ions are present in the primary ion beam used in the experiments as can be seen by comparing the cross sections displayed in Fig.~\ref{Fig:overview} panels a) and b) in the range of $1s \to 2p$ excitations.

Comparison of the ratios of experimental resonance strengths for the metastable and the ground-state components with the theoretical results obtained by Haso$\breve{\rm g}$lu {\textit{et al.}}~\cite{Hasoglu2010}, Wang and Zhou~\cite{Wang2007}, as well as Shi and Dong~\cite{Shi2009a} results in values for the fractional abundance of metastable C$^{+}(1s^22s2p^2~^4{\rm P})$ parent ions within 10$\pm$0.3\%.

Figure~\ref{Fig:1s2p} includes, as an example,  the results of the R-matrix calculations carried out by Haso$\breve{\rm g}$lu {\textit{et al.}}~\cite{Hasoglu2010} for both ion beam components. A weighted sum of the theoretical cross sections for 10\% metastable and 90\% ground-term components convoluted with a 38-meV FWHM Gaussian is compared with the experimental data. The theory data agree well with the experimental results for the dominant resonances. While peak areas are also predicted quite well for the smaller resonances there are shifts of up to 0.5~eV in the calculated resonance energies for the $^4{\rm P}^o$ and $^2{\rm S}$ K-vacancy states. Previous R-matrix calculations~\cite{Schlachter2004a} showed better agreement for the positions of the latter resonances but differed slightly in size and position for the strongest ($1s2s^22p^2~ ^2{\rm P}$) resonance. Further comparisons of the experimental data with the theoretical cross sections mentioned above will be shown and discussed below.

\begin{figure*}
\includegraphics[width=15.5cm]{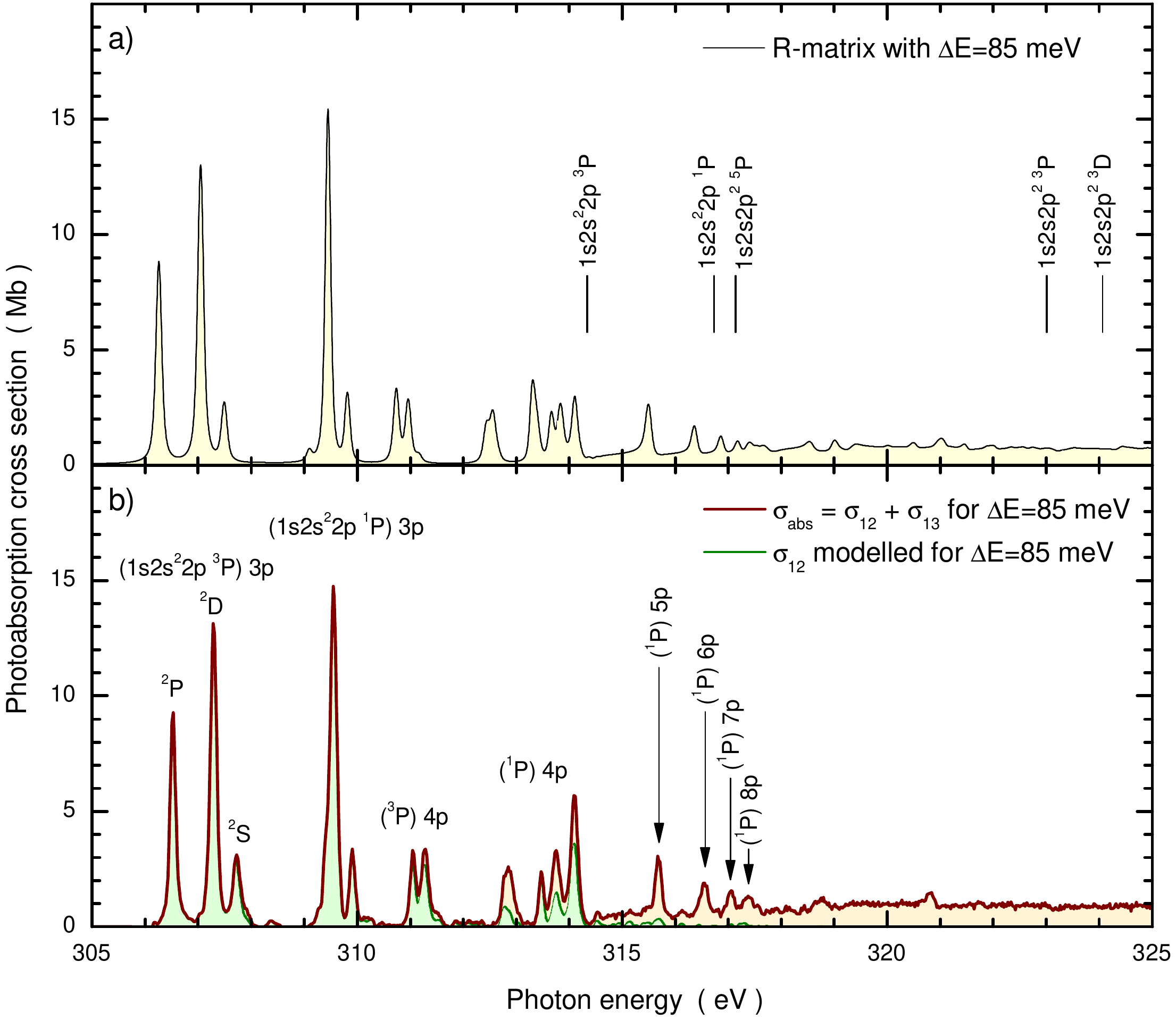}
\caption{\label{Fig:absorption} (color online) Calculated and measured cross sections of  C$^{+}$ ions interacting with photons in the energy range 305 - 325~eV. Panel a) shows the theoretical cross section for photoabsorption by ground-state C$^{+}$ ions~\cite{Hasoglu2010}. For comparison with the experimental results in panel b) the R-matrix results were convoluted with a 85-meV FWHM Gaussian. Energies of K-vacancy levels of C$^{2+}$ ions relative to the ground level of C$^{+}$ were calculated by using the Cowan code and are indicated by vertical bars. The associated K-edge terms are also indicated. Panel b) displays the experimentally derived cross section (see text) for single ionization of C$^{+}$ ions by single photons at 85~meV resolution represented by a thin (green) solid line with (light green) shading. The bold (brown) solid line with light (orange) shading represents the total experimental absorption cross section at 85~meV resolution. It is the sum of the experimental single and double photoionization cross sections at a photon beam bandwidth of 85~meV. Peak features in the spectrum have been identified by calculations using the Cowan code and, for the $1s2s^22p~^3{\textrm P}~3p$ manifold, by the assignments provided by Sun {\textit{et al.}}~\cite{Sun2011}.
}
\end{figure*}

The good agreement of the R-matrix photoabsorption calculations with the experimental cross sections for single ionization by single photons in the energy range 287 - 291~eV shows that photoabsorption in this energy range is dominated by K-shell excitations, $1s \to 2p$, of C$^+$ ions with subsequent single Auger decay populating the final C$^{2+}$ product ion channel. This situation changes at higher photon energies as Fig.~\ref{Fig:absorption} illustrates.

Figure~\ref{Fig:absorption} presents the comparison of theory and experiment at photon energies between 305 and 325~eV. Panel a) shows the photoabsorption cross section calculated by Haso$\breve{\rm g}$lu {\textit{et al.}}~\cite{Hasoglu2010} convoluted with a 85~meV FWHM Gaussian distribution function (solid line with light (yellow) shading). The lowest K-shell ionization thresholds relative to the $1s^2 2s^2 2p~^2{\rm P}$ ground-state term of C$^+$  are indicated by vertical bars. They were calculated by using the Cowan code~\cite{Cowan1981,Fontes2015}. Computations with different sets of configurations showed substantial configuration interaction in the initial C$^+$ and K-shell ionized C$^{2+}$ ions. Therefore, a relatively large set of configurations was used for the calculations of the K-shell ionization thresholds. For the initial C$^+$ ions the following configurations were included:
$1s^2 2s^2    2p$,
$1s^2  2p^3$,
$1s^2 2p 3s^2$,
$1s^2   2p  3p^2$,
$1s^2    2p  3d^2$,
$1s^2 2s^2  3p$, and
$1s 2s    2p^3$. For the K-shell ionized C$^{2+}$ ions the set of configurations comprised
$1s 2s^2    2p$,
$1s     2p^3$
$1s     2p 3s^2$,
$1s     2p  3p^2$,
$1s     2p   3d^2$,
$1s 2s^2     3p$, and
$2s    2p^3$. In spite of the relatively large calculation, the resulting K-edges may have uncertainties of the order of 1~eV. For example, the $1s 2s^2 2p~^1{\rm P}_1$ level which is the series limit of the $1s 2s^2 2p~(^1{\rm P}) np$ Rydberg resonance series, clearly seen in both the theoretical and experimental spectra,  should be near 317.6~eV rather than 316.7~eV as calculated with the Cowan code.

The GIPPER code~\cite{Fontes2015} predicts the cross section for direct C$^+$ K-shell ionization to be about 0.8~Mb just above threshold which is in fairly good agreement with the R-matrix calculation for photoabsorption and the experimental data for double ionization at energies beyond 319~eV. The K-edges associated with $^3{\rm P}_0$ (0.20~Mb), $^3{\rm P}_1$ (0.40~Mb) and $^1{\rm P}_1$ (0.20~Mb) constitute  almost the full K-shell photoabsorption  cross section resulting from the R-matrix calculations at energies above the $^1{\rm P}_1$ threshold.

Panel b) of Fig.~\ref{Fig:absorption} displays the experimental cross section for single ionization of C$^+$ by a single photon  as a thin (green) solid line with (light green) shading. There is remarkably good agreement of the experimental single-ionization cross section and the theoretical photoabsorption cross section at energies up to about 312~eV. With increasing photon energy  single-ionization appears to die out and there is no sign of the K edge. Understanding this phenomenon is straightforward. When a K-shell electron is removed from C$^+$ the dominating reaction to be expected is an Auger decay of the resulting K-vacancy state in C$^{2+}$ by which a further electron is removed and a C$^{3+}$ product ion is formed. Therefore, direct K-shell ionization results predominantly in net double ionization of the initial C$^+$ ion. Indeed, the associated cross section function shows clear evidence of K edges at energies above approximately 314.3~eV.

With a much smaller probability a double-Auger decay may follow the removal of a K-shell electron from C$^+$ producing C$^{4+}$ and thus contributing to net triple ionization. This contribution is very small as evidenced by the results shown in panel d) of Fig.~\ref{Fig:overview}. There is a clear signature of the K-edge with a steep rise of the triple-ionization cross section at energies above 316~eV, however, the ionization continuum is about a factor 100 smaller than that in double ionization. As will be demonstrated below, radiative stabilization of a K vacancy in C$^{2+}$ has a probability that is approximately two to three orders of magnitude smaller than that of Auger decay. Hence, photoabsorption by C$^+$ in the present energy range results almost exclusively in net single or double ionization. The sum of the associated cross sections $\sigma_{12}$ and $\sigma_{13}$, respectively, is a very good approximation for the total photoabsorption cross section. Therefore, panel b) of Fig.~\ref{Fig:absorption} includes that sum of the experimentally determined $\sigma_{12}$ and $\sigma_{13}$ at a photon energy bandwidth of 85~meV. It is represented by the bold solid (brown) line with (light orange) shading. Since single ionization was measured at 38~meV resolution and double ionization at 85~meV resolution, the experimental cross section $\sigma_{12}$ was convoluted with a 76.03-meV FWHM Gaussian to model a resolution of 85~meV = $(38^2 + 76.03^2)^{1/2}$~meV before the two cross sections were added.

The experimentally derived cross section $\sigma_{abs} = \sigma_{12} +\sigma_{13}$ is in very satisfying agreement with the theoretical photoabsorption cross section. Only details in the structure far beyond the K edge appear to be somewhat different with two pronounced features in the experimental spectrum  at about 318.7~eV and 320.8~eV which are not as obvious in the theoretical result. One has to keep in mind in this context, that the theory curve is for ground-state C$^+$ ions only, while the ion beam employed in the experiment included a 10\% fraction of ions in metastable $1s^2 2s 2p^2~^4{\rm P}$ levels. Apparently, this beam contamination does not have a strong influence on the measured cross sections at energies beyond 305~eV.

The two features at 318.7~eV and 320.8~eV mentioned above (see also Fig.~\ref{Fig:overview}, highest-energy resonance peaks in panel b) have not been identified. Most likely, they are associated with double excitations in which a K-shell electron and a L-shell electron are simultaneously promoted by one absorbed photon. Identification of the most prominent resonance features in the photoabsorption spectra is complicated by the very large number of levels that can potentially contribute. Again the GIPPER code~\cite{Fontes2015} was employed to see how much absorption oscillator strength is contributed by which levels at which excitation energies. Given the uncertainty of the calculated energies a level-by-level identification was not attempted.

For the calculations, a large set of configurations was included:
$1s^2 2s^2    2p$,
$1s^2  2p^3$,
$1s^2  2p  3s^2$,
$1s^2 2p 3p^2$,
$1s^2  2p 3d^2$,
$1s^2 2s^2 3p$,
$1s 2s 2p^2 np$,
$1s 2s 2p^2 ns$,
$1s 2s^2 2p np$,
$1s 2p^3 np$, with $n=2,3,...9$ and
$1s 2s 2p^2 nd$, with $n=3,4,...9$.
%$1s 2s    2p^3$,
%$1s 2p^4$,
%$1s 2s^2 2p 3p$,
%$1s 2p^3 3p$,
%$1s 2s^2 2p 4p$,
%$1s 2p^3 4p$,
%$1s 2s^2 2p 5p$,
%$1s 2p^3 5p$,
%$1s 2s^2 2p 6p$,
%$1s 2p^3 6p$,
%$1s 2s^2 2p 7p$,
%$1s 2p^3 7p$,
%$1s 2s^2 2p 8p$,
%$1s 2p^3 8p$,
%$1s 2s^2 2p 9p$,
%$1s 2p^3 9p$,
%$1s 2s 2p^2 3s$,
%$1s 2s 2p^2 4s$,
%$1s 2s 2p^2 5s$,
%$1s 2s 2p^2 6s$,
%$1s 2s 2p^2 7s$,
%$1s 2s 2p^2 8s$,
%$1s 2s 2p^2 9s$,
%$1s 2s 2p^2 3p$,
%$1s 2s 2p^2 4p$,
%$1s 2s 2p^2 5p$,
%$1s 2s 2p^2 6p$,
%$1s 2s 2p^2 7p$,
%$1s 2s 2p^2 8p$,
%$1s 2s 2p^2 9p$,
%$1s 2s 2p^2 3d$,
%$1s 2s 2p^2 4d$,
%$1s 2s 2p^2 5d$,
%$1s 2s 2p^2 6d$,
%$1s 2s 2p^2 7d$,
%$1s 2s 2p^2 8d$, and
%$1s 2s 2p^2 9d$.
More than 9000 transitions were considered and their dipole oscillator strengths $f_a$ for absorption calculated. Because of the uncertainties of the calculated resonance energies the relative positions of K-shell excited levels and the magnitude of the oscillator strengths were used to assign configurations to the most prominent peak features in the spectrum. According to this analysis the energy range 306 - 308~eV is dominated by levels in the $(1s2s^2 2p~^3{\rm P})~3p$ manifold, the range 309 - 310.5~eV by levels in the $(1s2s^2 2p~^1{\rm P})~3p$ manifold, and the range 310.5 - 312~eV appears to be due to $(1s2s^2 2p~^3{\rm P})~4p$ levels. The Rydberg series of $(1s2s^2 2p~^3{\rm P})~np$ resonances converges to the $1s2s^2 2p~^3{\rm P}$ K-shell ionization edge at about 314.5~eV and is buried under resonances associated with $(1s2s^2 2p~^1{\rm P})~4p$ levels. The $(1s2s^2 2p~^1{\rm P})~np$ Rydberg sequence with $n = 5, 6, 7, 8,...$ dominates in the energy range 315.5 - 317.8~eV.

Figures~\ref{Fig:1s2p} and \ref{Fig:absorption} demonstrate that the K-shell excited levels of C$^+$ in the energy range up to about 312~eV predominantly decay by emission of one electron. These levels are associated with $1s \to 2p$, $1s \to 3p$, and partly with $1s \to 4p$ excitations. When the principal quantum number of the outermost excited electron increases further, the probability for the emission of two electrons increases and the final product is a C$^{3+}$ ion. At energies beyond the $1s 2s^2 2p~^3{\rm P}$ K-shell ionization threshold even the resonantly excited C$^+$ ions decay almost exclusively by two-electron emission. As a result, the cross section for absorption of a photon by a C$^+$ ion in that energy range is nearly identical with the cross section for net double ionization.

 \begin{figure}
\includegraphics[width=\columnwidth]{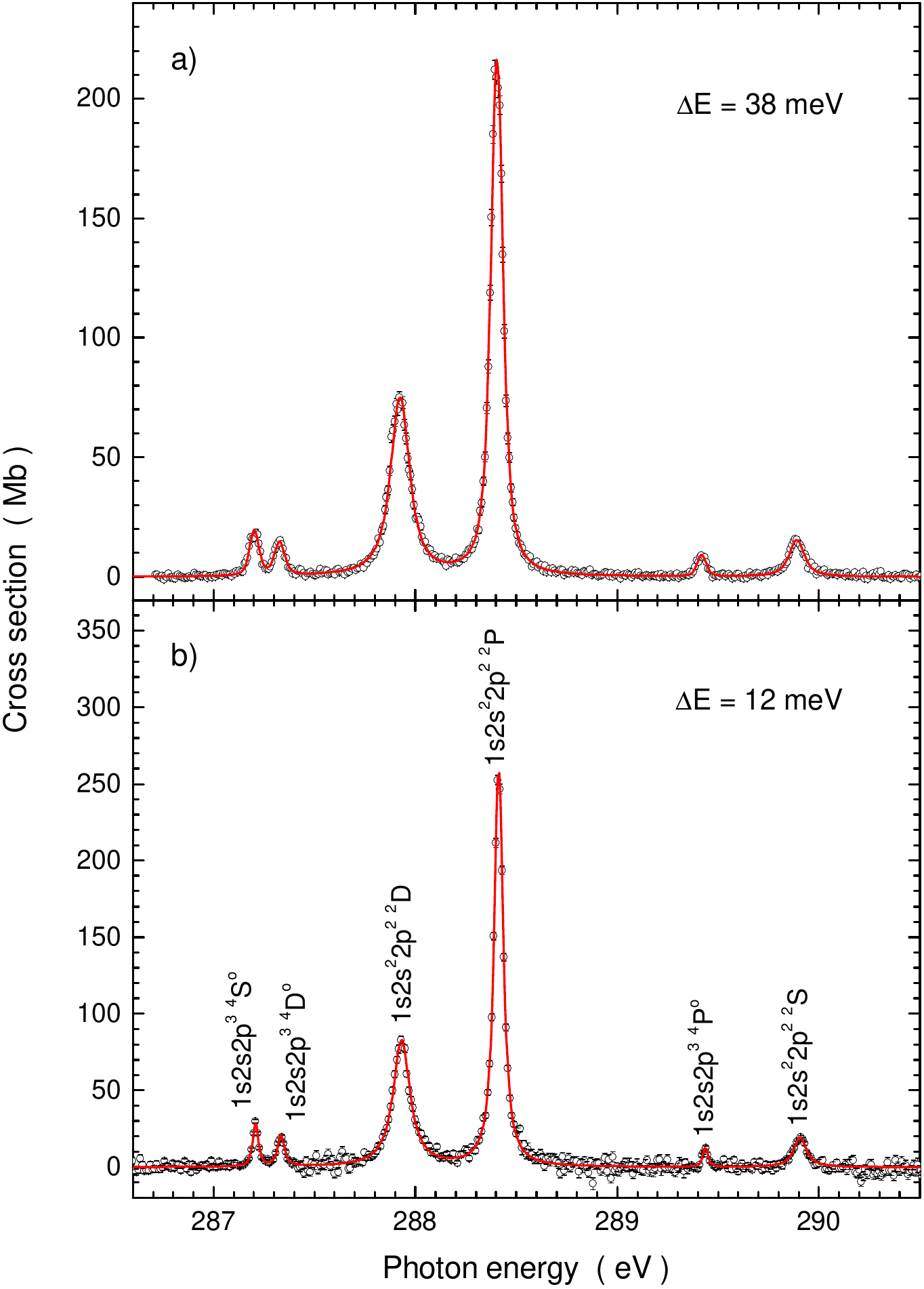}
\caption{\label{Fig:hires} (color online) Measured cross sections for net single ionization of C$^{+}$ ions measured at resolution 38~meV (panel a) and 12~meV (panel b). The parent ion beam contained 10\% metastable $^4$P and 90\% $^2$P ground-term ions. The solid (red) lines are fits to the two spectra as described in the main text. }
\end{figure}

The present experiment with its low detector background and its high photon flux provided much better conditions than those faced in the experiments of Schlachter {\textit{et al.}}~\cite{Schlachter2004a} on K-shell excitation followed by single-Auger decay at the Advanced Light Source (ALS) in Berkeley.  Hence, it was possible to increase the resolving power by approximately four times that of the previous experiment. The smallest photon energy bandwidth reached by Schlachter {\textit{et al.}} was 50~meV. At this bandwidth only the two strongest resonances in the spectrum could be measured with moderate statistics. Compared to the previous best resolution the bandwidth was reduced in the present experiment to about one fourth. The complete spectrum of $1s \to 2p$ resonances of C$^+$ ions in the energy range 286.5 - 290.5~eV was measured at a resolution of 12~meV. The resulting high-resolution cross section for single ionization of C$^+$ ions is shown in Fig.~\ref{Fig:hires} panel b) together with a 38-meV resolution measurement in panel a) in which excellent statistical precision was reached.

\begin{table*}
\caption{\label{tab:resonanceparameters} Parameters characterizing the $1s \to 2p$ excitation resonances  associated with ground-state C$^+(1s^2 2s^2 2p~^2{\rm P})$ (first three columns) and metastable C$^+(1s^2 2s 2p^2~^4{\rm P})$  parent ions (last three columns). The top entry of each column is the K-vacancy term excited by a single photon. The resonance energies $E_{\rm res}$ are given with three decimals because their relative uncertainties are of the order of only 5~meV. The absolute calibration uncertainty is 30~meV. Throughout this paper  the numbers in round brackets following a measured quantity denote the uncertainty of the last digits.
The following parameters are provided as extracted from resonance fits to the measured spectra:
$E_{\rm res}$, the resonance energy relative to the initial term energy;
$\Gamma_{\rm L}$, the Lorentzian width of the excited level;
$S_{1e}$, the partial resonance strength for net single ionization;
$S_{2e}$, the partial resonance strength for net double ionization;
$S_{3e}$, the partial resonance strength for net triple ionization;
$\lambda$, the transition wavelength;
$\tau$, the lifetime of the excited level;
$S$, the total resonance strength for photoabsorption;
$f_{ik}$, the oscillator strength for absorption from level $|i\rangle$ to level $|k\rangle$;
$A_{ki}$; the associated Einstein coefficient for radiative decay of the excited level;
$\Gamma_\gamma$, the radiative  (partial) width of the excited level;
$\Gamma_{1e}$, the partial width for the emission of one electron;
$\Gamma_{2e}$, the partial width for the emission of two electrons;
$\Gamma_{3e}$, the partial width for the emission of three electrons;
$A_{1e}$, the single-Auger decay rate;
$A_{2e}$, the double-Auger decay rate;
$A_{3e}$, the triple-Auger decay rate. Quantities ($S_{1e}$, $S_{2e}$, $S_{3e}$, and $S$) that depend on the fraction of ions in the specified initial term are normalized to a fraction of 100\% so that the derived parameters (such as $f_{ik}$) retain their intended meaning.}
\begin{ruledtabular}
\begin{tabular}{lllllll}
parameter & $1s2s^22p^2~^2{\rm D}$    &  $1s2s^22p^2~^2{\rm P}$  &  $1s2s^22p^2~^2{\rm S}$  & $1s2s2p^3~^4{\rm S}^o$  & $1s2s2p^3~^4{\rm D}^o$ & $1s2s2p^3~^4{\rm P}^o$   \\

\hline
$E_{\rm res}$ [eV] & 287.931(30) & 288.413(30)  & 289.906(30) & 287.208(30)  & 287.334(30) & 289.437(30)  \\
$\Gamma_{\rm L}$ [meV]  &  101(3)  & 49(2)  & 70(5)  & 24(3)  & 37(4)  & 22(5)  \\
$S_{1e}$ [Mb\,eV]  &  14.6(2.2) & 23.8(3.6) & 2.48(40) & 13.4(4.0)  & 12.6(3.8) & 5.46(1.7) \\
$S_{2e}$ [Mb\,eV]  &  0.38(6)  & 0.81(13) & 0.084(13) & 0.36(11) & 0.34(10) & 0.15(5) \\
$S_{3e}$ [Mb\,eV]  &  0.0019(10) & 0.0032(16) &  & &  &  \\$\lambda$ [nm]     & 4.30604(45) & 4.29884(45)  & 4.27670(44) & 4.31688(45) & 4.31499(45) & 4.28363(44) \\$\tau$ [fs] & 6.5(2)  &  13.4(5)  & 9.4(7)  & 27.4(3.4)  & 17.8(2.4)  & 29.9(6.8)  \\
$S$ [Mb\,eV]  &  15.0(4.1) & 24.9(7.0)  &  2.57(91) & 14.3(7.7) & 13.0(7.1) & 5.65(3.4) \\
$f_{ik}$  &  0.137(38) & 0.227(64)  & 0.0234(83) & 0.130(71)  & 0.118(65) & 0.051(31) \\
$A_{ki}$ [10$^{11}$~s$^{-1}$] & 2.95(81)  & 8.18(2.3)  & 2.56(91)  & 14.0(7.6) & 2.54(1.4)  & 1.9(1.1)  \\
$\Gamma_\gamma$ [meV]  &  0.19(6)  & 0.54(15)  & 0.17(6)  & 0.92(50)  & 0.17(9)  & 0.12(7)  \\
$\Gamma_{1e}$ [meV]  &  98.3(10.0)  & 46.9(5.1)  & 67.6(8.3)  & 22.5(3.6)  & 35.9(6.0)  & 21.3(5.3)  \\
$\Gamma_{2e}$ [meV]  &  2.5(3)  & 1.6(2)  & 2.3(3)  & 0.61(10)  & 0.95(16)  &  0.60(15) \\
$\Gamma_{3e}$ [meV]  &  0.013(7)  & 0.0063(32)  &   &   &   &   \\
$A_{1e}$ [10$^{13}$~s$^{-1}$] & 14.9(1.6)  & 7.1(8)  & 10.3(1.3)  &  3.4(6) & 5.5(1.0)  & 3.2(8) \\
$A_{2e}$ [10$^{11}$~s$^{-1}$] & 38.5(4.0)  & 24.4(2.6)  & 34.6(4.3)  &  9.2(1.5) & 14.5(2.4)  & 9.1(2.3) \\
$A_{3e}$ [10$^{9}$~s$^{-1}$] & 19(10)  & 9.6(5.0)  &   &   &   &  \\

\end{tabular}
\end{ruledtabular}

\end{table*}

From the experimental results displayed in Fig.~\ref{Fig:hires} resonance parameters for all the six peaks related to $1s \to 2p$ excitations can be extracted using a global fit to both spectra. However, one has to consider that all the K-shell excited levels are multiplets as are the initial electronic levels of the parent C$^+$ ions. Due to the natural widths of the excited levels, none of these multiplets can be resolved in an experiment, even at infinite resolving power. The electron-cyclotron-resonance ion source used in the present experiments is known to produce a plasma with high electron temperatures. Therefore, it is easily possible to ionize and excite ions in a wide range of charge states. Excited states of ions with sufficiently long lifetimes can survive the time between their population in the plasma and the extraction of the ions from the source. They can further survive the time-of-flight between the source and the photon-ion interaction region which was about 30~$\mu$s in the present case of C$^+$  ions.
Candidates for such survival are the first four excited electronic (fine-structure) levels associated with the $1s^2 2s^2 2p~^2{\rm P^o}$ ground term and the energetically lowest metastable $1s^2 2s 2p^2~^4{\rm P}$ term.

The NIST Atomic Spectra Database~\cite{NIST2015} provides the following energies for the relevant levels populated by the ions in the parent ion beam:
0.00000~eV $(^2{\rm P^o}_{1/2})$, 0.007863~eV $(^2{\rm P^o}_{3/2})$, 5.33173~eV $(^4{\rm P}_{1/2})$, 5.33446~eV $(^4{\rm P}_{3/2})$, and 5.33797~eV $(^4{\rm P}_{5/2})$. So the possible initial terms feature fine-structure splitting by up to almost 8~meV. The $1s \to 2p$ excited terms in Fig. ~\ref{Fig:hires} are $1s 2s^2 2p^2~^2{\rm D}$,$^2{\rm P}$,$^2{\rm S}$ and $1s 2s 2p^3~^4{\rm S^o}$, $^4{\rm D^o}$, $^4{\rm P^o}$. According to Shi and Dong~\cite{Shi2009a} the fine-structure splitting within the $^2{\rm D}$ term is 12~meV and 13~meV within the $^2{\rm P}$ term. Similar splittings are found for the K-shell excited quartet terms.

Selection rules for electric dipole transitions between the initial and final terms limit the number of possible transitions between the initial and final levels. Nevertheless, the 6 peaks observed in the experiment in the energy range 287 - 291~eV comprise 27  significantly contributing level-to-level transitions~\cite{Shi2009a}. The $^2{\rm D}$ resonance populated from the ground term is made up of 3 transitions with transition energies differing by up to 19~meV. The $^2{\rm P}$ resonance comprises 4 transitions separated by a maximum of 21~meV and the $^2{\rm S}$ peak consists of 2 transitions separated by 8~meV. Similarly, the 3 peaks in the quartet system populated by excitation of ions initially in one of the $1s^2 2s 2p^2~^4{\rm P}$ levels are composed of 18 levels in total. The $^4{\rm S^o}$ resonance consists of 3 lines separated by a maximum of 3~meV. The $^4{\rm D^o}$ peak contains 8 lines separated by a maximum of 14~meV and the $^4{\rm P^o}$ peak comprises 7 transitions separated by a maximum of 9~meV. All the fine-structure splittings given above were taken from the theoretical results obtained by Shi and Dong~\cite{Shi2009a}.

The multiplet structure complicates the interpretation of linewidths obtained by theory and experiment. In previous work addressing K-shell excitation of B-like ions C$^+$~\cite{Schlachter2004a}, N$^{2+}$~\cite{Gharaibeh2014} and O$^{3+}$~\cite{McLaughlin2014} the multiplet nature of the resonances was not considered. Not accounting for the fine-structure effects is acceptable as long as the level splitting is much smaller than the natural or lifetime width $\Gamma$ of the individual transitions. This has to be investigated for each case. Without further discussion of fine-structure splitting the Lorentzian width $\Gamma_{\rm L}$ extracted from resonance fits has to be interpreted as an effective natural width which is not necessarily identical with the lifetime width.

In order to quantify the possible effects of fine-structure splittings on the widths of the $1s \to 2p$ photoexcitation resonances the information on resonance energies, natural line widths and oscillator strengths provided by Shi and Dong~\cite{Shi2009a} for each of the 27 possible $1s \to 2p$ resonance transition was used to construct a suitable model for fitting the experimental photoionization cross section. By careful inspection of table 3 of their publication it became apparent that they had entered wrong numbers for the oscillator strengths for the three transitions $^4{\rm P^o}_{5/2} \to~^4{\textrm S}_{3/2}$, $^4{\textrm P^o}_{3/2} \to~^4{\textrm S}_{3/2}$, and $^4{\rm P^o}_{1/2} \to~^4{\textrm S}_{3/2}$. The corrected numbers (provided in private communication between C.-Z. Dong and A.M.) are 0.577, 0.404, and 0.186, respectively. The $gf$ values calculated in length form were divided by the statistical weights of the $^4{\textrm P^o}$ initial levels and then the resonance strengths were calculated for each combination of initial levels $|i\rangle$ to final levels $|k\rangle$ from the formula  {(see Appendix)}
\begin{equation}
\label{totalstrength}
S = \int_{E_\gamma}  \sigma_{tot}(E_\gamma) d{E_\gamma} = f_{ik} \times 109.761~{\rm Mb\,eV}
\end{equation}
where $E_\gamma$ is the photon energy and $\sigma_{tot}$ the cross section for total photoabsorption via the $|i\rangle \to |k\rangle$ transition. The quantity  $f_{ik}$ is the oscillator strength of that transition.

The six apparent peak features were then fitted by 27 Voigt profiles, each of which is characterized by a Lorentzian and a Gaussian width, the resonance energy, and the resonance strength. In addition, a constant background cross section was allowed. Rather than using a 109-parameter fit, a number of constraints were introduced. The Gaussian widths characterizing the experimental energy spread  are all the same for each of the 27 profiles within one measured spectrum. Furthermore, the Lorentzian widths associated with each of the fine-structure levels of one term have been assumed to be identical. Only the six energies of the lowest-energy transitions within each of the six apparent resonance features were used as fit parameters while the remaining transition energies were fixed by the fine-structure splittings calculated by Shi and Dong~\cite{Shi2009a}. Also, only six parameters were used to fit the 27 resonance strengths by keeping the calculated relative resonance strengths within each peak feature identical to those predicted by theory~\cite{Shi2009a}.  With all these constraints, the Gaussian width, six Lorentzian widths, six apparent resonance energies and six factors for correcting the theoretical resonance strengths were extracted from the experiment.

\begin{figure}
\includegraphics[width=\columnwidth]{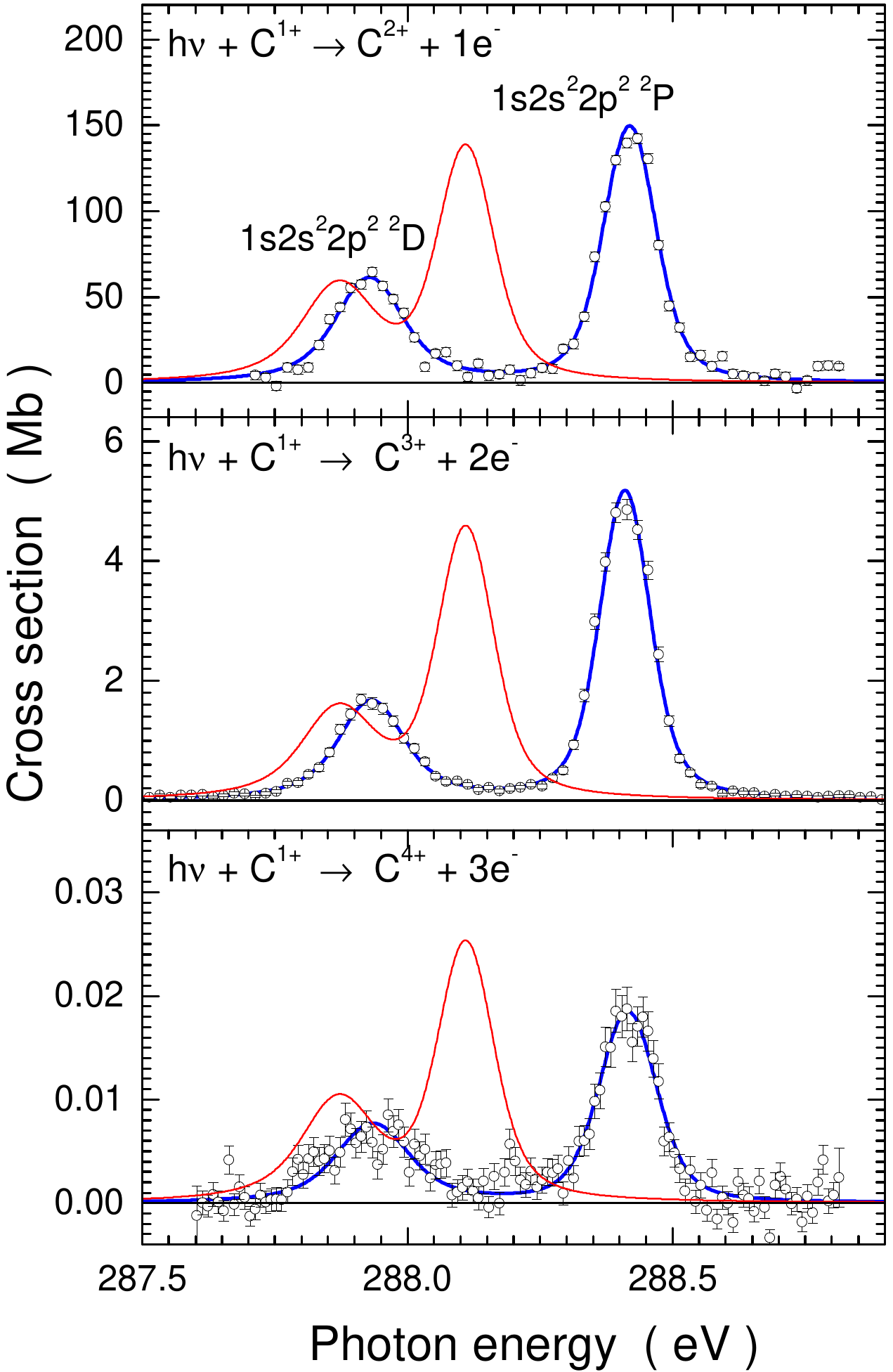}
\caption{\label{Fig:multPI} (color online) Cross section for single, double and triple ionization of ground-state C$^{+}$ ions by single photons at 92~meV bandwidth. The resonances seen in all the observed channels are associated with K-shell-excited C$^{+}(1s2s^22p^2~^2{\textrm D},~^2{\textrm P})$ terms. The dark solid (blue) lines represent the fits to the experimental spectra with Voigt profiles from which Auger decay probabilities were obtained. The light (red) solid lines show theory-based absolute cross sections derived from calculations of single-, double- and triple-Auger decay rates by Zhou {\textit{et al.}}~\cite{Zhou2016}. For details see main text.
}
\end{figure}

The resulting parameters for the six $1s \to 2p$ excitation resonances are provided in Table~\ref{tab:resonanceparameters}. The first row shows the excitation energies found for the six resonances. Their absolute calibration uncertainty is $\pm 30$~meV. Yet the energies are given with three decimals because the relative uncertainty is less than 5~meV.  The second and third rows show the remaining parameters resulting from the fit of the spectra in Fig.~\ref{Fig:hires}, i.e.,  the natural line widths $\Gamma = \Gamma_{\rm L}$ and the resonance strengths $S_{1e}$ for one-electron removal. The estimated uncertainties are given by the numbers in brackets. In similar fits to experimental spectra (see Figs.~\ref{Fig:overview} and ~\ref{Fig:multPI}) the resonance strengths $S_{2e}$ for two-electron removal, and $S_{3e}$ for three-electron removal were also determined.
The measured strengths were normalized to 100\% considering the relative fractions of ions in the ground term (90\%) and the metastable term (10\%) in the measurement. In other words, the resonance strengths obtained from the measurements were divided by 0.9 in case of the resonances populated from the ground-state term and by 0.1 in the case of resonances populated from the metastable parent ion term.

The remaining parameters provided in Table~\ref{tab:resonanceparameters} have all been derived from the observables given in the first five rows. The relevant formulas are collected in the appendix which describes the procedure used for extracting all the additional information about the observed resonances. For convenience, the sixth row of Table~\ref{tab:resonanceparameters} shows the wavelength of the transitions observed in the experiment. These are readily obtained from Eq.~\ref{Eq:lambdaik}. The lifetimes of the core-excited terms are easily calculated from Eq.~\ref{Eq:tau}. The determination of the resonant-absorption strength $S$ is more involved and requires additional information about radiative transition rates. An iterative procedure which converges after two steps provides $S$ and the oscillator strength $f_{ik}$ as well as the radiative decay rate $A_{ki}$ and the partial width $\Gamma_\gamma$ for radiative decay of the core-excited term. Following from these parameters and the measured resonance strengths $S_{ne}$ ($n=1,2,3$) for the removal of $n$ electrons, the transition rates $A_{ne}$ for single-, double-, and triple-Auger decay as well as the partial widths of these decays can also be obtained.

\begin{figure}
\includegraphics[width=\columnwidth]{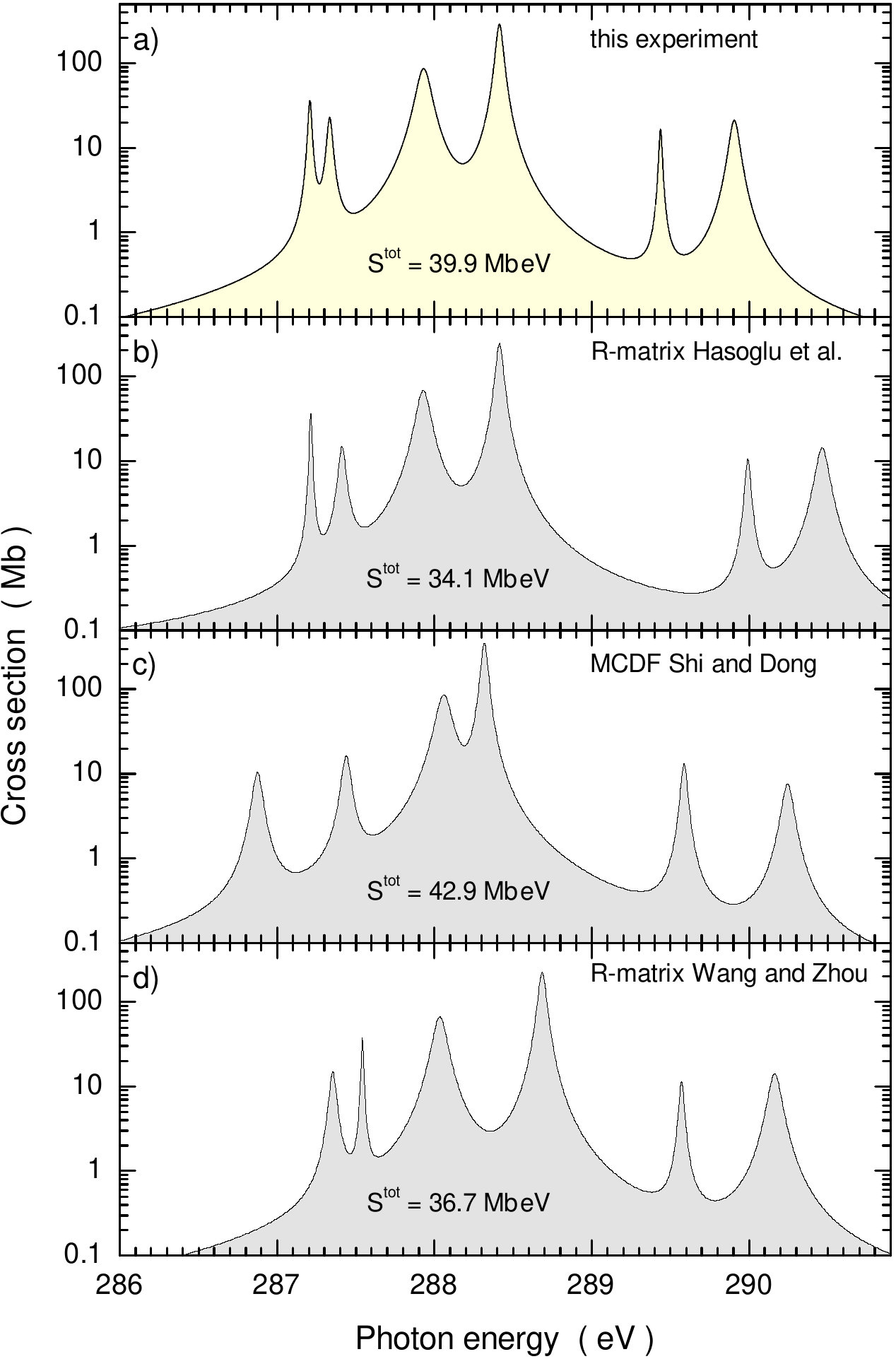}
\caption{\label{Fig:1sto2p-absorption} (color online) Natural-line-width absorption cross sections of C$^{+}$ determined by the present experiment with a 90\% fraction of $^2{\textrm P}$ ground-level and 10\% of $^4{\textrm P}$ metastable ions in the parent ion beam. The experimental spectrum (panel a) is compared with calculations by Haso$\breve{\rm g}$lu {\textit{et al.}}~\cite{Hasoglu2010} (panel b), Shi and Dong~\cite{Shi2009a} (panel c) and Wang and Zhou~\cite{Wang2007} (panel d). The theory spectra were modeled assuming the experimental distribution of ground-term and metastable ions. Each panel displays the total resonance strength $S^{tot}$ contained in the associated photoabsorption spectrum.
}
\end{figure}

The information collected in Table~\ref{tab:resonanceparameters} can be used for constructing the natural (unconvoluted) absorption cross section related to $1s \to 2p$ excitation of C$^+$. This can be directly compared to the results of theoretical calculations. In principle, for such comparisons the cross sections for exclusively ground-term C$^+$ ions can be separated from the cross sections for exclusively metastable C$^+$ ions. Instead, comparisons are made assuming the conditions of the present experiment, i.e., 10\% metastable and 90\% ground-level ions. The experimentally derived natural cross section for such a mixture is shown in panel a) of Fig.~\ref{Fig:1sto2p-absorption}.

The simulated experimental cross section is compared with the results of two R-matrix calculations~\cite{Hasoglu2010,Wang2007} and the cross section reconstructed from the detailed information provided by Shi and Dong on the basis of their multiconfiguration Dirac-Fock calculations~\cite{Shi2009a}. All cross sections are shown on a logarithmic scale to emphasize the smaller resonances in the comparison. The calculations show spectra which are quite similar to the experimental result. Obvious differences are mainly in the positions of the six resonance features. However, one has to keep in mind that the deviations are typically only within several tens of meV and reach no more than about 550~meV in a few cases. Such deviations at level energies of approximately 290 eV are less than 0.19\%.

In Fig.~\ref{Fig:1sto2p-absorption}, the total resonance strength of each spectrum is provided. The average of the theoretical resonance strengths is 37.9~Mb\,eV which is only 9.5\% from the experimental value. The maximum deviation is 14.5\% which is still within the experimental uncertainty. While the resonance energies and the peak areas in the theory spectra are close to the experiment there are dramatic differences in the peak widths between different theoretical approaches and between theory and experiment. Quantitative comparisons are provided in Table~\ref{tab:parametersandtheory} where selected parameters listed in Table~\ref{tab:resonanceparameters} are compared with theoretical and experimental data available in the literature.

\begin{table*}
\caption{\label{tab:parametersandtheory} Selected parameters from Table~\ref{tab:resonanceparameters} in comparison with experimental and theoretical results for the $1s \to 2p$ excitation resonances  associated with ground-state C($1s^2 2s^2 2p~^2{\rm P}$) (first three columns) and metastable C($1s^2 2s 2p~^4{\rm P}$)  parent ions (last three columns). The peak assignment by Schlachter {\textit{et al.}}~\cite{Schlachter2004a} as well as that by Wang and Zhou~\cite{Wang2007} has been corrected according to the discussion by Haso$\breve{\rm g}$lu {\textit{et al.}}~\cite{Hasoglu2010} and the results of Shi and Dong~\cite{Shi2009a}. The calculations by Shi and Dong are differential in the total angular momentum and appropriately weighted sums were entered into this table.}
\begin{ruledtabular}
\begin{tabular}{lllllll}
 parameter & $1s2s^22p^2~^2{\rm D}$    &  $1s2s^22p^2~^2{\rm P}$  &  $1s2s^22p^2~^2{\rm S}$  & $1s2s2p^3~^4{\rm S}^o$  & $1s2s2p^3~^4{\rm D}^o$ & $1s2s2p^3~^4{\rm P}^o$   \\

\hline
$E_{\rm res}$ [eV] \\
experiment\\
this work & 287.931(30) & 288.413(30)  & 289.906(30) & 287.208(30)  & 287.334(30) & 289.437(30)  \\
\cite{Schlachter2004a} & 287.93(3)   & 288.40(3)  &  289.90(3) &    & 287.25(3)  &  289.42(3)  \\
\cite{Jannitti1993a}  & 287.91(7) & 288.59(7) & 289.95(7) & 287.09(7) & 288.02(7) & 289.80(7) \\

theory\\
\cite{Hasoglu2010} spectrum fit  & 287.929 & 288.414 & 290.465  & 287.213  & 287.413   & 289.992    \\
\cite{Hasoglu2010} their Table I  & 287.966 & 288.303 & 290.279  & 286.997 & 287.465   & 290.012    \\
\cite{Shi2009a} & 288.061  & 288.317  & 290.240  & 286.875  & 287.440  &  289.587   \\
\cite{Wang2007} & 288.034  & 288.685  & 290.163  & 287.354  & 287.542  & 289.571    \\
\cite{Sun2011} & 287.90    & 288.56   & 289.92   &          & 287.33   & 289.63   \\
\cite{Schlachter2004a} & 287.96  & 288.63  &  289.97 & 287.29  & 287.73  &  289.46  \\
\hline

$f_{ik}$ \\
this work  &  0.137(38) & 0.227(64)  & 0.023(8) & 0.13(7)  & 0.12(7) & 0.051(31) \\
theory\\
\cite{Hasoglu2010} spectrum fit & 0.100  & 0.191  & 0.020  &  0.084  & 0.095   &   0.053   \\
\cite{Shi2009a} length form & 0.139  & 0.249  & 0.028  & 0.002*  & 0.132  &  0.079   \\
\cite{Shi2009a} velocity form & 0.126  & 0.226  & 0.025  & 0.002*  & 0.125  &  0.074   \\
\cite{Shi2009a} revised & & & & 0.0973* & & \\
\cite{Wang2007} length form & 0.1015  & 0.1927  & 0.0197  & 0.0968  &  0.0844 &  0.0547   \\
\cite{Wang2007} velocity form & 0.0977  & 0.1855  & 0.0194  & 0.0931  &  0.0812 &  0.0534   \\

\hline

$\Gamma$ [meV]  \\
experiment\\
this work &  101(3)  & 49(2)  & 70(5)  & 24(3)  & 37(5)  & 22(5)  \\
\cite{Schlachter2004a} & 105(15)   & 59(6)   & 112(25)   &     & 110(40)   &  55(25)   \\
theory\\
\cite{Hasoglu2010} spectrum fit & 94  & 50  &  87 & 16   & 47   &  36    \\
\cite{Shi2009a} & 101  & 43  & 77  & 57  & 67  &  42   \\
\cite{Wang2007} & 96  & 54  & 88  & 47  & 16  &  35   \\
\cite{Schlachter2004a} & 103   & 62   & 93   & 25   & 84   &  52   \\

\hline

$A_{ki}$ [10$^{11}$~s$^{-1}$]  \\
experiment\\
this work &  2.95(81)  & 8.18(2.3)  & 2.56(91)  & 14.0(7.6)  & 2.5(1.4)  & 1.9(1.1)  \\
theory, length form\\
\cite{Sun2011} & 2.25 & 6.98  & 2.24  &    & 3.22   &  1.91    \\

\hline

$A_{1e}$ [10$^{13}$~s$^{-1}$]  \\
experiment\\
this work & 14.9(1.6) & 7.12(77)  & 10.3(1.3)  & 3.42(55)  & 5.45(92)  & 3.23(80)  \\
theory, length form\\
%\cite{Sun2011} & 32.5 & 14.6  & 18.9  &    &    &      \\
\cite{Chen1988d} & 13.1 & 7.22 & 12.6 & 8.90 & 7.25 & 5.67 \\
\cite{Hasoglu2010} & 13.9 & 7.49 & 13.1 & 2.35 & 6.99 & 5.41 \\
\cite{Hasoglu2006} & 17.6 & 7.88 & 15.5 & 3.13 & 9.75 & 7.40 \\
\cite{Shi2009a} & 15.4 & 6.44 & 11.7 &  10.1 & 8.53 & 6.29\\

\end{tabular}
\end{ruledtabular}
* Entries in table III of reference~\cite{Shi2009a} are most likely misprinted; the corrected number, 0.0973, was calculated by Shi and Dong (private communication with A.M. 2017) using the FAC code.
\end{table*}

The parameters for which comparisons are possible are the resonance energies $E_{\rm res}$, the oscillator strengths $f_{ik}$, the natural width $\Gamma$ which is equal to the Lorentzian width $\Gamma_L$ found in the fits, the radiative decay rate $A_{ki}$ and the single-Auger decay rate $A_{1e}$.

The resonance energies found in the present investigation agree with most of the other experimental findings within their error bars. Deviations are observed for the positions of the smaller peaks in the spectrum particularly when comparing with the spectroscopic work of Jannitti {\textit{et al.}}~\cite{Jannitti1993a}. Differences with theoretical calculations have already been discussed and are below a level of $1.90 \times 10^{-3}$.

The experimentally derived oscillator strengths for absorption have estimated uncertainties between 28\% for the strongest ground-term resonances and about 60\% for the resonances associated with metastable parent ions. The reason for the relatively large error bars is in the uncertainty of the 10\% fraction of the metastable component in the parent ion beam. With only few exceptions the theoretical oscillator strengths are well within the experimental error bars.

With an instrumental width of only 12~meV the determination of natural line widths for core-excited levels of  C$^+$ ions can be expected to be quite accurate. Compared to the previous experiment by Schlachter {\textit{et al.}}~\cite{Schlachter2004a} the uncertainties of some of the $\Gamma$ values could be substantially reduced by factors up to 8. Nevertheless the width of the broadest resonance, associated with the core-excited $^2$D term, is within 4\% in both experiments. In other cases the uncertainties quoted in the earlier experiment were slightly too optimistic. The theoretical natural widths show substantial scatter by factors of over 4. On average theory predicts broader resonances for metastable parent ions than what the experiment yields.

The radiative decay rates derived from the present experiment can be directly compared with the calculations by Sun {\textit{et al.}}~\cite{Sun2011} who used the saddle-point variation and complex-rotation methods including relativistic and mass polarization corrections in first-order perturbation theory. The calculated radiative transition rates are in agreement with the experiment within the experimental uncertainties. Finally, the total transition rates for single-Auger decay found by the present measurement are compared with the results of four different theoretical calculations. For the resonances associated with ground-term parent ions, theory and experiment are in quite satisfying agreement. However, in few cases there are deviations exceeding a factor of 2.

The comparisons show that the accurate calculation of atomic transition parameters for a relatively simple atomic ion with only 5 electrons is still a challenge. With the results obtained in the present experiment for double- and triple-Auger decays the challenge becomes even greater. New theoretical attempts are being developed to deal with multi-electron ejection subsequent to inner-shell excitation (see e.g.~\cite{Schippers2016a}). However, so far only one calculation has been published addressing the multi-electron, and particularly the triple-Auger decay processes associated with K-shell excitations of C$^+$ ions that have been studied in the present experiment. But very encouraging new theoretical work that includes calculations also of direct triple-Auger decay is underway~\cite{Liu2017}.

The last part of this paper is devoted to the quantitative description of multiple ionization of C$^+$ ions as the result of the absorption of a single photon. For the strongest resonances found in single ionization, additional measurements of single, double and triple ionization were carried out at a fixed energy resolution of 92~meV~\cite{Mueller2015a}. The lowest panel in Fig.~\ref{Fig:multPI} shows the unambiguous observation of C$^{4+}$ product ions arising from K-shell excited C$^{+}(1s2s^22p^2~^2{\textrm D},~^2{\textrm P})$ states which are populated from the ground-state term of C$^{+}$ via one-photon absorption $\gamma + {\textrm C}^{+}(1s^22s^22p~^2{\textrm P}) \to {\textrm C}^{+}(1s2s^22p^2~ ^2{\textrm D},~^2{\textrm P})$. The resulting K-vacancy states can obviously decay by the emission of up to three electrons. Other conceivable explanations for the observation of C$^{4+}$ product ions such as two-photon absorption or one-photon absorption plus one or more interactions with residual-gas particles were discussed in our previous Letter~\cite{Mueller2015a} and shown to be negligible.

\begin{table}
\caption{\label{tab:comparison} Comparison of the present resonance parameters for C$^+(1s^2 2s^2 2p~^2{\textrm P}) \to {\textrm C}^+(1s2s^22p^2~^2{\textrm D},~^2{\textrm P})$  excitations with calculations by Zhou {\textit{et al.}}~\cite{Zhou2016} including the triple-Auger decay rate. The data obtained by Zhou {\textit{et al.}} were taken from their Table I for $A_{1e}$ and derived from the ratios given in their Table IV.
}
\begin{ruledtabular}
\begin{tabular}{lcc}
 parameter &  $1s2s^22p^2~^2{\rm D}$   & $1s2s^22p^2~^2{\rm P}$     \\

\hline
$E_{\rm res}$  [eV], this work & 287.931(30) & 288.413(30)   \\
$E_{\rm res}$ [eV], Zhou {\textit{et al.}} & 287.87 & 288.11   \\
\hline
$\Gamma$ [meV], this work &  101(3)  & 49(2)  \\
$\Gamma$ [meV], Zhou {\textit{et al.}} & 114  & 57  \\
\hline
$A_{1e}$ [10$^{13}$~s$^{-1}$], this work & 14.9(1.6)  & 7.1(8) \\
$A_{1e}$ [10$^{13}$~s$^{-1}$], Zhou {\textit{et al.}} & 17.4 & 8.66 \\
$A_{2e}$ [10$^{11}$~s$^{-1}$], this work & 38.5(4.0)  & 24.4(2.6) \\
$A_{2e}$ [10$^{11}$~s$^{-1}$], Zhou {\textit{et al.}} & 53.4 & 27.1 \\
$A_{3e}$ [10$^{9}$~s$^{-1}$], this work  & 19(10) & 9.6(5.0) \\
$A_{3e}$ [10$^{9}$~s$^{-1}$], Zhou {\textit{et al.}} & 31.2  & 15.8  \\

%$\Gamma_{1e}/\Gamma$ [\%], this work & 97.2 $\pm$ 1 & 96.7 $\pm$ 1\\
%$\Gamma_{1e}/\Gamma$ [\%], Zhou {\textit{et al.}} & 97.0  & 96.9  \\
%\hline
%$\Gamma_{2e}/\Gamma$ [\%], this work & 2.6 $\pm$ 0.4 & 3.3 $\pm$ 0.5 \\
%$\Gamma_{2e}/\Gamma$ [\%], Zhou {\textit{et al.}} & 2.979  & 3.039 \\
%\hline
%$\Gamma_{3e}/\Gamma$ [\%], this work & 0.013 $\pm$ 0.007  &  0.013 $\pm$ 0.007 \\
%$\Gamma_{3e}/\Gamma$ [\%], Zhou {\textit{et al.}} & 0.0174  & 0.0177 \\

\end{tabular}
\end{ruledtabular}

\end{table}

Along with the experimental data, Fig.~\ref{Fig:multPI} presents theoretical cross section curves constructed from calculations of single-, double- and triple-Auger decay of K-shell excited C$^{+}(1s2s^22p^2~^2{\rm D},~^2{\rm P})$ terms. Decay rates for Auger processes starting from these terms with emission of one, two or three electrons were computed by Zhou {\textit{et al.}}~\cite{Zhou2016} and are compared in Table~\ref{tab:comparison} with the experimentally derived decay rates. From the theoretical decay rates as well as the calculated resonance energies $E_{\rm res}$ and natural widths $\Gamma$, also provided in Table~\ref{tab:comparison}, relative theoretical cross sections can be constructed for single, double, and triple ionization of ${\rm C}^{1+}(1s^22s^22p~ ^2{\rm P})$ ions by a single photon via excitation of K-shell excited C$^{+}(1s2s^22p^2~^2{\rm D},~^2{\rm P})$ terms. For producing absolute cross sections additional information on the absorption oscillator strength  is required which has not been provided by Zhou {\textit{et al.}}. By using the experimentally derived values of $f_{ik}$ (see Table~\ref{tab:parametersandtheory}) plus the parameters calculated by  Zhou {\textit{et al.}} as presented in Table~\ref{tab:comparison} the light (red) solid curves in Fig.~\ref{Fig:multPI} have been computed which represent the absolute cross sections for single (upper panel), double (middle panel) and triple (lower panel) ionization of C$^+$. For direct comparison with the experiment the theoretical cross sections were convoluted with 92-meV FWHM Gaussians.

The resonance energies obtained by Zhou {\textit{et al.}} are slightly off, however, the differences are at most 300~meV corresponding to 0.1\% of the associated resonance energies. The magnitudes of the cross sections, partly being adjusted to the experimental data by using the experimentally derived oscillator strengths, give remarkable agreement for all three final charge states of the ionized ions which requires realistic multi-electron-ejection decay rates. The comparison of these rates in Table~\ref{tab:comparison} shows a level of agreement that is very encouraging for further developments in this new field of research.

\section{Summary}
In this paper, which is a follow-up of a previous Letter~\cite{Mueller2015a}, details of the experiments are presented. In particular, the experimental data are compared to available theoretical results from different computational approaches. From high-resolution absolute cross-section measurements at a resolving power of up to 24\,000, exceeding that of previous experiments by a factor of four, complete sets of transition parameters for $1s \to 2p$ excitations of ground-term and metastable C$^+$ ions could be derived. These include not only the resonance energies, natural widths and resonance strengths but also the transition rates for radiative decay as well as single-, double-, and triple-Auger decays. Oscillator strengths, partial decay widths, transition wavelengths, and lifetimes of K-shell excited terms are also readily computed from the extracted information. Comparison of these quantities with data available in the literature show mixed agreement. This is remarkable since C$^+$ with only five electrons is a relatively simple system. It is all the more encouraging that now also the decay rates for multi-electron ejection up to the triple-Auger decay can be theoretically accessed producing results that are in remarkable agreement with the present measurements opening a new window into the important field of many-electron processes in atomic systems.

\begin{acknowledgments}
This research was carried out in part at the light source PETRA III at DESY, a member of the Helmholtz Association (HGF).
We gratefully acknowledge support from Bundesministerium f\"{u}r Bildung und Forschung provided within the "Verbundforschung" funding scheme (contract numbers 05K10RG1,  05K10GUB, 05K16RG1, 05K16GUC). We thank the P04 beamline team for their support of our work. We are grateful to M. F. Haso$\breve{\rm g}$lu and collaborators for providing their numerical R-matrix data. We thank C.-Z. Dong and Y.-L. Shi for providing corrected theory data for the absorption oscillator strength of metastable carbon ions. M.M. is thankful for the financial support via DFG SFB925/A3. S.K. acknowledges support from the European Cluster of Advanced Laser Light Sources (EUCALL) project which has received funding from the \textit{European Union's Horizon 2020 research and innovation programme} under grant agreement No. 654220. A.M. acknowledges support from Deutsche Forschungsgemeinschaft under project number Mu 1068/22.

\end{acknowledgments}

\appendix*
\section{Determination of atomic transition rates and related quantities from absolute high-resolution cross-section measurements of photoionization resonances}

{This appendix introduces quantities often used in atomic and molecular spectroscopy and provides information on their mutual relations. Such information can be found in textbooks  and scientific papers (see for example \cite{Cowan1981,Thorne1988,Martin2006,Wiese2009}). However, the use of different definitions such as line strength and resonance strength may cause confusion. Moreover, different unit systems in theoretical treatments and in the resulting equations provide additional pitfalls. It is felt therefore, that a consistent set of spectroscopic quantities and equations should be introduced and discussed in the present context. In all formulas SI units are used. On the basis of the  relations provided in the following, many of the data presented in this paper have been derived. }

Photoionization of an atom or an ion can proceed in a direct process in which an electron is removed from one of the occupied subshells. Alternatively, an inner-shell electron can be excited to an autoionizing state that relaxes by Auger decay. The latter process is characterized by a resonance in the cross section which is associated with the intermediate multiply excited state. This state has an intrinsic width, the natural width $\Gamma$, and is found at the photon energy $E_{res}$. This energy is required to populate the intermediate excited level from the initial state which can be the ground level or a long-lived excited level. Depending on the experimental resolution and the natural widths of the intermediate excited levels, individual resonances can be separated in the measured photoionization spectrum, i.e., in the cross section for the production of ions in a given charge state as a function of photon energy.

With sufficient experimental resolution the natural width of each individual resonance can be determined. From the absolute measurement of the photoionization cross section $\sigma$ the resonance strength, which is essentially the peak area under the resonance,
\begin{equation}
\label{resonancestrength}
S = \int_{E_\gamma}  \sigma(E_\gamma) d{E_\gamma}
\end{equation}
is obtained. $E_\gamma$ is the photon energy. {This definition of $S$ follows the experiment-oriented approach pursued, for example, in the treatment by Thorne~\cite{Thorne1988} who defines an integrated absorption coefficient $k$ which is related to the absorption oscillator strength $f_{ik}$ via
\begin{equation}
\label{absorptionstrength}
\int_{\nu}  k(\nu) d{\nu} = \frac{e^2}{4 \epsilon_0 m_e c} n_1 f_{ik}
\end{equation}
where $\nu$ is the frequency of the absorbed light, $e$ the elementary charge, $\epsilon_0$ the electric constant, $m_e$ the electron rest mass, $c$ the vacuum speed of light, and $n_i$ the number density of absorbing atoms in state $|i\rangle$ per unit volume. A beam of photons passing the distance $x$ through the absorbing dilute medium is attenuated by the factor $\exp(-k x) = \exp(-\sigma n_1 x)$. By replacing $k(\nu)$ in Eq.~\ref{absorptionstrength} with $\sigma n_1$ and multiplying both sides of the equation with Planck's constant $h$ and dividing by $n_1$, one obtains
\begin{equation}
\label{Sversusfik}
S = \int_{E_\gamma}  \sigma(E_\gamma) d{E_\gamma} = \frac{h e^2}{4 \epsilon_0 m_e c} f_{ik} = 109.761 \textrm{Mb eV} f_{ik}.
\end{equation}
Since $S$ follows from measured cross sections it is possible to derive the absorption oscillator strength from the experiment.}

Eq.~\ref{absorptionstrength} is valid under the assumption that quantum mechanical interferences between resonant and nonresonant as well as between neighboring resonances can be neglected.  As in the present experiment, the intermediate core-excited level may undergo different Auger processes and cascades of Auger decays may be possible leading to different final charge states of the initial atom or ion. Each channel is described by a cross section $\sigma_{ne}$ with $n$ characterizing the number of electrons removed from the initial atom or ion. The intermediate excited level can also decay by the emission of electromagnetic radiation. The associated channel has not been observed in  merged-beam photon-ion experiments so far. Yet it is possible to extract information on the rate for radiative transitions. As an example, photoexcitation of ground-term C$^+(1s^2 2s^2 2p~^2{\textrm P})$ ions to K-shell-excited C$^+(1s 2s^2 2p^2~^2{\textrm D})$ and C$^+(1s 2s^2 2p^2~^2{\textrm P})$ is discussed below. Absolute cross sections $\sigma_{1e}$, $\sigma_{2e}$, and $\sigma_{3e}$ for net single, double and triple photoionization, respectively, are included in Fig.~\ref{Fig:overview} and separately displayed  in Fig.~\ref{Fig:multPI}.

From a detailed high-resolution experiment, like the present one, the resonance energy $E_{res}$ and the natural width $\Gamma$ can be obtained as well as the resonance strength $S_{ne}$ for each of the $n$-electron-removal channels
\begin{equation}
\label{Eq:resstrength-ne}
S_{ne} = \int_{E_\gamma}  \sigma_{ne}(E_\gamma) d{E_\gamma} .
\end{equation}
The number $n$ of electrons finally removed from the parent atom or ion depends on the atomic number, the charge state,  the level structure and on the core-hole state populated in the photoabsorption process. For the present examples at most $n=3$ electrons were removed. The cross section for net triple photoionization is already very small (by a factor of $\approx 10^{-4}$ compared to single photoionization).

The transition wavelength $\lambda_{ik}$ follows immediately as
\begin{equation}
\label{Eq:lambdaik}
\lambda_{ik} = hc/E_{res} = 1 239. 841 9739~{\rm nm}\,{\rm eV} / E_{res}
\end{equation}
 (the 2014 CODATA values for fundamental physical constants are used throughout this paper). The lifetime $\tau$ of the intermediate resonant level is {
\begin{equation}
\label{Eq:tau}
\tau = \hbar/ \Gamma = 6.582119514 \times 10^{-16}~{\rm eV\,s} / \Gamma
\end{equation}
with $\hbar$, as usual, equal to Planck's constant divided by $2\pi$.}

Next, the absorption oscillator strength $f_{ik}$ follows from the absorption resonance strength $S$. This quantity is the sum of the individual resonance strengths of all decay channels
\begin{equation}
\label{Eq:totabsstrength}
S = S_{1e} + S_{2e} + S_{3e} + S_\gamma .
\end{equation}
So far, the resonance strength contained in the photon-emission channels is not known and photoemission is not observed. However, the absorption oscillator strength $f_{ik}$ for the transition from the initial level $|i\rangle$ to the intermediate autoionizing level $|k\rangle$ is directly related to the Einstein coefficient $A_{ki}$ which describes the rate of radiative decay from level $|k\rangle$ to level $|i\rangle$. The oscillator strength follows from the relation  {(see Eq.~\ref{Sversusfik})
\begin{equation} \label{Eq:oscistrength}
f_{ik} = S \frac{4 \epsilon_0 m_e c}{h e^2} = S \times 9.1107 \times 10^{-3}/~{\rm Mb\,eV}.
\end{equation}
}
The oscillator strength $f_{ik}$ for an electric dipole transition $i \to k$  can also be expressed by its relation with the transition rate $A_{ki}$~\cite{Wiese2009}
\begin{equation}
f_{ik} = 1.4992 \times 10^{-14}~{\rm s/nm^2} \times {\lambda}_{ki}^2 \frac{g_k}{g_i} A_{ki}  ,
\end{equation}
where  $g_k$ and $g_i$  are the statistical weights of $|k\rangle$ and $|i\rangle$, respectively. The sum of all transition rates for radiative decays to final levels $|f\rangle$  provides the partial width ${\Gamma}_\gamma$ of the resonant intermediate level
\begin{equation}
{\Gamma}_\gamma = \hbar  \sum_f A_{kf}  .
\end{equation}
For the present examples of K-shell resonance transitions in C$^+$ the only relevant radiative-decay paths of the intermediate C$^+(1s 2s^2 2p^2~^2{\textrm D})$ and C$^+(1s 2s^2 2p^2~^2{\textrm P})$ levels lead back to the
C$^+(1s^2 2s^2 2p~^2{\textrm P})$ initial term so that
\begin{equation}
{\Gamma}_\gamma = \hbar  A_{ki} .
\end{equation}
With the equations given above and the expectation that radiative decay is much less probable than Auger decay, $S$ and $f_{ik}$ can be iteratively determined.

In the first approximation, one can set
\begin{equation}
\label{Eq:SAuger}
S^{(1)} = S_{1e} + S_{2e} + S_{3e}
\end{equation}
and derive
\begin{equation} \label{Eq:oscifirst}
f^{(1)}_{ik} = S^{(1)} \times 9.1107 \times 10^{-3}/~{\rm Mb\,eV}
\end{equation}
and hence a first approximation of the radiative width of the intermediate excited state is given by
\begin{equation}
{\Gamma}^{(1)}_\gamma = \hbar  A^{(1)}_{ki} = \hbar  f^{(1)}_{ik} \frac{g_i}{g_k} /(1.4992 \times 10^{-14}~{\rm s/nm^2} \times {\lambda}_{ki}^2).
\end{equation}
In addition, a  first approximation of the resonance strength for radiative decay of the intermediate excited state can be derived as
\begin{equation} \label{Eq:oscifirst}
S^{(1)}_\gamma = S^{(1)} \frac{{\Gamma}^{(1)}_\gamma}{\Gamma}.
\end{equation}

For the examples under discussion $S^{(1)}_\gamma$ is approximately  $2 \times 10^{-3}$ to $1.1 \times 10^{-2}$ of $S^{(1)}$, i.e., the correction of $S^{(1)}$ is at least an order of magnitude smaller than the experimental uncertainty of $S^{(1)}$. It is therefore justified to set
\begin{equation}
\label{Eq:SAuger}
S = S^{(2)} = S^{(1)} + S^{(1)}_\gamma.
\end{equation}
With $S$ now known, $f_{ik}$ follows from Eq.~\ref{Eq:oscistrength} and then
\begin{equation}
A_{ki} =  f_{ik} \frac{g_i}{g_k} /(1.4992 \times 10^{-14}~{\rm s/nm^2} \times {\lambda}_{ki}^2) .
\end{equation}

The partial widths for $n$-electron removal are readily obtained from
\begin{equation}
\label{Eq:Gamne}
\Gamma_{ne} = \frac{S_{ne}}{S} \Gamma
\end{equation}
and the Auger decay rates for the emission of $n$ electrons are
\begin{equation}
\label{Eq:Gamne}
A_{ne} = \Gamma_{ne} / \hbar.
\end{equation}
All these quantities have been derived from the measurement of $1s \to 2p$ excitations of C$^+$ and are listed in Table~I. As far as available, they are compared with the literature in Table~II.

%\bibliography{k3AM}

%

\end{document}